\documentclass[emulateapj,epsfig]{article}
\usepackage[onecolumn]{emulateapj}
\usepackage[]{epsfig}
\usepackage{graphicx}
\def\perc{\%\,\,}
\hoffset -1.5truecm \lefthead{Menci et al.}
\righthead{}
\def\newline{\hfil\break}
\begin{document}
\title{The Red Sequence of High-Redshift Clusters: a Comparison with Cosmological Galaxy Formation Models}
\author{N. Menci$^1$, P. Rosati$^2$, R. Gobat$^2$, V. Strazzullo$^3$, A. Rettura$^4$, S. Mei$^{5,6}$, R. Demarco$^4$ }
\affil{$^1$INAF - Osservatorio Astronomico di Roma, via di Frascati 33, I-00040 Monteporzio, Italy}
\affil{$^2$European Southern Observatory, Karl-Schwarzschild-Strasse 2, D-85748 Garching, Germany}
\affil{$^3$National Radio Astronomy Observatory, P.O. Box O, Socorro, NM 87801}
\affil{$^4$Department of Physics and Astronomy, Johns Hopkins University, Baltimore, MD 21218}
\affil{$^5$University of Paris Denis Diderot,  75205 Paris Cedex 13, France}
\affil{$^6$GEPI, Observatoire de Paris, Section de Meudon, 92195 Meudon Cedex, France}
\smallskip

\begin{abstract} We compare the results from a state-of-the-art
    semi-analytic model of galaxy formation with spectro-photometric 
    observations of distant galaxy clusters observed in the range
    $0.8\leq z\leq 1.3$. We investigate the properties of their red
    sequence (RS) galaxies and compare them with those of the field at the same 
    redshift. In our model we find that i) a
    well-defined, narrow RS is obtained already by $z\approx 1.2$;
    this is found to be more populated than the field RS, analogously to what
    observed and predicted at $z=0$; ii) the predicted U-V rest-frame 
    colors and scatter of the cluster RS at $z=1.2$ have average values 
    of 1 and 0.15 respectively, with a 
    cluster-to-cluster variance of $\approx 0.2$ and $\approx 0.06$, respectively. 
    The scatter of the RS of cluster
    galaxies is $\approx $ 5 times smaller than the corresponding field value;
    iii) when the RS galaxies are considered, the mass growth
    histories of field and cluster galaxies at $z\approx1.2$ are
    similar, with 90 \perc of the stellar mass of RS galaxies at
    $z=1.2$ already formed at cosmic times $t=2.5$ Gyr, and 50 \perc
    at $t=1$ Gyr; v) the predicted distribution of stellar ages of RS
    galaxies at $z=1.2$ peaks at $3.7$ Gyr for
    both cluster and field populations; however, for the latter the distribution
    is significantly skewed toward lower ages. When compared with
    observations, the above findings show an overall consistency,
    although the average value $\Delta_{U-V}\approx 0.07$ of the observed cluster
    RS scatter at $z\approx1.2$ is smaller than the corresponding model
    central value. We discuss the physical origin and the significance
    of the above results in the framework of cosmological galaxy
    formation.
\end{abstract}


\section{Introduction}

Environment-dependent properties of galaxies constitute a natural
test-ground for cosmological theories of galaxy formation. These
envisage the properties of the galaxy populations to originate from the
primordial density field dominated by the Dark Matter (DM); the 
collapse and growth of small-scale
perturbations (leading to galaxy progenitors) is thus modulated by the
underlying large-scale density field whose overdense regions will
collapse to form groups and clusters of galaxies.

Among the galaxy properties, the color distribution constitutes a
major issue, as colors measure the ratio of the present star
formation rate to the overall mass of formed stars in a given
galaxy. Three fundamental observational results deal with the star
formation history and the color distribution of field galaxies: a) The
global star formation decrease from $z=1.5$ to the present by
almost 2 dex (see, e.g., the compilation by Hopkins 2004 and references therein); 
b) The downsizing (Cowie et al. 1996), 
i.e., a stronger ($\sim 3$ dex) and
earlier (starting at least at $z\approx 2$) decline of star formation
in massive galaxies compared  to a slower ($\approx$ 1 dex)
decline of the star formation rate of the present blue/late-type
galaxies from $z\sim 1.5$; correspondingly, an inverse correlation of
the age of stellar population with the galaxy stellar mass is
observed; c) The bimodality (Strateva et al. 2001; Baldry et al. 2004;
Bell et al. 2004), i.e., a marked segregation in color
between the blue galaxy population (dominant at faint magnitudes and
for low-mass galaxies) and the red population (dominant for bright,
massive galaxies) populating the ''red sequence'' (RS).

In hierarchical clustering models, the first property naturally
results from the high star formation rate at high redshifts $z\gtrsim
2$ sustained by the high cooling efficiency in the dense
environments of galaxy progenitors and by effective starbursts; the
decline at $z\lesssim 2$ follows from the exhaustion of the gas reservoir
converted into stars at higher $z$. The interpretation of downsizing
in hierarchical galaxy formation is also straightforward: bright,
massive galaxies form from the coalescence of protogalaxies which
collapsed in a biased, overdense region of the primordial density
field; thus such progenitors collapsed and began to form stars at
earlier cosmic epochs, and their cold gas reservoir is exhausted
earlier thus yielding older (and hence redder) stellar population in
the final galaxy. The origin of the bimodality is more controversial;
recent works (see discussion in Neistein, van der Bosch, Dekel 2006;
Dekel \& Birnobim 2006; Menci et al. 2006) suggest that
self-regulation of star formation (due to Supernovae feedback) can be
effective below a critical (DM) mass scale $M\approx 5\times
10^{11}\,M_{\odot}$; the inclusion of feedback from AGNs then quenches 
further star formation in massive haloes.

The above properties are strongly modulated by the galaxy
environment. Not only the morphology has long been recognized to
strongly depend on the environment (with the fraction of early-type
galaxies increasing in groups and clusters of galaxies, Spitzer \& Baade
1951), but also the fraction of galaxies belonging to the blue
population has been recently recognized to decrease in dense
environments (Baldry et al. 2004, 2006; Gerke et al. 2007; Cooper et
al. 2006). The environmental dependence of the blue fraction
$f_b$ is stronger than the above mentioned luminosity dependence
observed in the field. This means that the environmental density
enhances the probability for an early star formation followed by
quenching even for low or intermediate mass galaxies.

The scenario emerging from the above observational framework is the
following: galaxies flow from the blue population to the red sequence
(corresponding to the decline of star formation at $z<2$) at rates and
cosmic epochs that depend very strongly on their mass and 
environment. Massive galaxies have converted most of their gas into
stars and moved to the red sequence very early at $z>2$ in both
cluster and field, since at high redshift their star formation was not
effectively self-limited by feedback. On the other hand, lower mass
galaxies $M<10^{12}\,M_{\odot}$ move to the RS later in the
field compared to clusters. The speed-up of the transition to the red
sequence of intermediate/low- mass galaxies in dense environments
(groups and clusters) can be contributed by a number of physical
processes: i) Biased galaxy formation; in principle, galaxies later
included in groups and clusters form from clumps which collapsed in biased,
overdense regions of the primordial field, hence characterized by an
earlier star formation. However, the initial overdensity corresponding
to the final group or cluster may be so tiny that such an effect might
be too small.  ii) Starbursts resulting from galaxy merging and
fly-by (sometimes referred to as ''harassment'');  
while merging is actually enhanced in groups compared to
clusters (see Cavaliere, Colafrancesco \& Menci 1992), the minor - but
more frequent - starbursts originated from galaxy grazing encounters
may significantly accelerate the star formation at high redshift in
groups and clusters, and hence anticipate the transition to the red
sequence. Note however that this process is effective mainly for
mid-size galaxies which strike the best compromise between cross
section and abundance.  iii) Strangulation; in small-mass galaxies the
gas may be stripped off when they are included in the cluster (Larson,
Tinsley \& Caldwell 1980); in most semi-analytic models this is
captured by dispersing their hot gas mass through the whole host halo
(see, e.g., Balogh, Navarro \& Morris 2000). Note that our model, 
similarly to canonical semi-analytic approaches,  does not 
deal with the ram pressure stripping of the dense, cold gas in satellite galaxies. 
iv) AGN feedback; 
to some extent, such a mechanism must be at work, and there is a general
agreement (Ciotti \& Ostriker 1997; Silk \& Rees 1998, Haehnelt,
Natarajan \& Rees 1998, Fabian 1999) that it may contribute to
suppress any residual star formation in massive galaxies; in models 
where the AGN phase is triggered by galaxy interactions, the AGN feedback 
depends on the environment which modulates the encounter probability.
The effectiveness of the AGN feedback in quenching the black hole growth
and the subsequent star formation has been confirmed by recent aimed
simulations of galaxy collisions triggering AGN activity (Di Matteo,
Springel \& Hernquist 2005).  We have investigated the role of such a
process in a previous paper (Menci et al. 2006) showing that it can
affect the color distribution of most massive galaxies by removing the
residual fraction of such galaxies that otherwise would populate the
blue branch of the color distribution.

The relative role of all these processes is still matter of
investigation. Their dependence on
the mass-scale of the host structure (group or galaxy cluster) would provide
an important clue to disentangle their contribution to the observed 
environment-dependent properties of galaxies.  
Several recent papers (Cooper et al. 2007, Gerke et
al. 2007) point out that the difference in the properties of the RS 
between field and overdense environments remains unchanged up
to $z\approx 1$ but tends to vanish at redshifts $z\gtrsim 1.3$. Since
these results are based on samples which probe the environments up to
the group scale (rich clusters are not included), the epoch
$1.2\lesssim z\lesssim 1.5$ would represent a measure of the effectiveness of
poor/medium environments in modulating the star formation of member
galaxies; at such redshifts, the environment-induced decline of the
star formation would become stronger than the average cosmological
decline. Probing the transition to the RS of galaxies in
richer, extreme environment at $1\lesssim z\lesssim 1.5$ is then
crucial.

In this context, recent HST/ACS observation of a sample of clusters at
$0.8<z<1.3$ have fueled much progress. E.g., Blakeslee et
al. (2003, 2006) and Mei at al. (2006a; 2006b) have found a well
defined and extremely narrow red sequence with an intrinsic scatter of 
$\sim 0.05$ mag (in the color $i_{775}-z_{850}$ as observed  
with HST/ACS), much smaller than the corresponding field RS. This points
toward the existence of a strong environment-dependent effect which,
for galaxies in extremely dense environments, has already been
effective at $z\approx 1.5$, thus supporting the view that the denser
the environment, the earlier the transition to the RS.

How do such results compare with hierarchical models ? While such models
essentially capture the transition of faint galaxies to the RS 
(at least in the redshift range $z\leq 0.8$, see De Lucia et al. 2007; 
 for results of  hydro/N-body simulations see Romeo et al. 2008 ), 
the tightness of the RS constitutes an impressive constraint for them, 
since in the hierarchical scenario the growth of DM haloes constitutes 
- on average - a gradual process, driven by the merging of
sub-clumps, the progressive cooling of gas at the center followed by a
quiescent star formation. Since the tightness of the
observed RS indicates a fast transition of 
the galaxies to the same region of the color distribution acting on
very short time scales in the early epochs, this is particularly challenging
for hierarchical models which predict clusters of galaxies to be the last
structures to virialize (typically at $z<2 $). On the other hand, in
such models, high-redshift clusters also constitute the most biased
environments for galaxies to form, hence enhancing the effectiveness of
the processes i)-iv) described above in accelerating the transition to
the RS. Thus, comparing the observed properties of the RS in
high-z clusters with model predictions constitutes an extremely
sensitive probe of hierarchical models and of the physical processes there
included.

Here we compare the results of a state-of-the-art semi-analytic model
of galaxy formation with the properties of the RS of galaxy
clusters at high redshift, the most extreme (biased) galaxy
environments observed so far. We also emphasize that the observations
used in this work are currently the most suited for such an
analysis, as with ACS observations in the $i$ and $z$ bands one
can achieve the photometric accuracy to reliably estimate color
scatters out to $z\sim\! 1.3$.

The model we adopt is well suited for such an investigation since it
includes all the processes i)-iv) discussed above, which are though to
contribute to the environmental dependence of the star formation
histories. In particular, it includes both the burst of star
formation due to grazing encounters (important in the early building
up of massive galaxies) and a description of the AGN evolution which
has been extensively tested against observations both in the optical
(Menci et al. 2003) and in X-rays (Menci et al. 2004b).  In previous
papers (Menci et al. 2005; 2006) 
we have successfully compared the model with observations
concerning the galaxy color distribution at low redshifts as a
function of both luminosity and environment, and at high-redshift we
showed how the inclusion of AGN feedback allows to suppress
residual star formation in massive galaxies yielding the observed
abundance of EROs.

The paper is organized as follows: in Sects. 2 and 3 we briefly recall
the main properties of the semi-analytic model we use, and the
observational data set we compare with. In Sect. 4.1 we show the model
results concerning the color-magnitude relation both at low and at high 
redshift in the cluster and in the field, and we briefly discuss how it
compares with existing observations. In sect. 4.2 we focus on the
properties of the RS in high-redshift clusters, and compare the model results
concerning the scatter and the normalization of the RS with the ACS
observations of distant clusters. We also provide predictions
concerning the distribution of the scatter and the normalization of
the RS among the cluster population, which could be verified with
existing or future observations. In Sect. 4.3 we also compare the mass
assembly history and the age distribution of $z\approx 1.2$ galaxies in
the field and in clusters resulting from the model with estimates derived
from fitting observed galaxy colors and spectra with spectral
synthesis models. In Sect. 4.4 we interpret our results on the basis of
the physical processes which determine the scatter of the RS in 
hierarchical models. In Sect. 5 we draw our conclusions.

\section{The Model}

Here we recall the key properties of the semi-analytic model we adopt,
referring to our previous papers for details. The DM 
haloes of protogalaxies collapse from overdense regions of the
primordial DM density field, taken to be a Gaussian random density
field with a Cold Dark Matter (CDM) power spectrum in a ''concordance''
cosmology with $\Omega_{\Lambda}=0.7$,
$\Omega_{0}=0.3$, baryonic density $\Omega_b=0.04$, and Hubble constant
(in units of 100 km/s/Mpc) $h=0.7$. The normalization of the spectrum,
in terms of the variance of the field smoothed over a region of 8
$h^{-1}$ Mpc, is taken to be $\sigma_8=0.9$. Since the rms value of the
density field decreases with mass, progressively larger regions of the
density field collapse with increasing time (eventually leading to the
formation of groups and clusters of galaxies), and include the
previously formed condensations. The corresponding merging rate of the
DM haloes is provided by the Extended Press \& Schechter formalism (see
Bond et al. 1991; Lacey \& Cole 1993) and implemented as described in
detail in Menci et al. (2005, 2006).  The clumps included into larger
DM haloes may survive as satellites, merge to form larger galaxies due
to binary aggregations among satellites, or coalesce into the central
dominant galaxy due to dynamical friction. These processes take place
over time scales which increase over cosmic times, so the number of
satellite galaxies increases as the DM host haloes grow from groups to
clusters. All the above processes are implemented using the usual 
prescriptions of semi-analytic models as described in Menci et
al. (2005; 2006).

We adopt the treatment of the baryonic processes described in our
previous papers (Menci et al. 2005) with the same choice of free
parameters. These include the cooling at the center of galaxies, the
settling of the cooled gas with mass $m_c$ into a rotationally
supported disk with radius $r_d$ and circular velocity $v_d$, the
gradual (quiescent) conversion of such gas into stars at a rate $\dot
m_*$, and the stellar feedback which results in redistributing part of
the cooled gas to the hot gas phase (with mass $m_h$) at the virial
temperature of the halo. An additional channel for star formation
implemented in the model is provided by interaction-driven starbursts,
triggered not only by merging events but also by fly-by events. Such a
star formation mode provides an important contribution to the early
formation of stars in massive galaxies, as described in detail in
Menci et al. (2004a, 2005).

The model also includes a treatment of the growth of supermassive
black holes at the center of galaxies by interaction-triggered inflow
of cold gas, following the physical model in Cavaliere \& Vittorini
(2000). The accretion episodes, corresponding to the active AGN phases,
also trigger the AGN feedback onto the surrounding cold gas, whose
physical treatment is derived from Lapi, Cavaliere, Menci (2005). A
detailed description and testing of such a section of the model is
presented in Menci et al. (2006).

Thus, all processes which are candidate for enhancing the early star
formation in galaxies in dense environments (and for the subsequent
quenching of star formation) are included in the model.  The
distribution of gas mass, disk sizes, stellar masses, luminosities,
have been checked against observations in previous papers, as well as
the growth of AGNs and the corresponding redshift-dependent luminosity
functions both in the optical and in X-rays.

\section{The Data Set}

We compare our model mainly with the results from the HST/ACS
Intermediate Redshift Cluster Survey, which consisted of multi-color
observations of eight clusters with redshifts in the interval
$0.8\leq z\leq 1.3$. We focus on the observations of the RS of these
clusters, as described in Blakeslee et al. (2003, 2006), Homeier et
al.  (2006), Mei et al. (2006a,b, 2008). We also use the analysis by
Gobat et al. (2008) and Rettura et al. (2008) of the 10-band spectral 
energy distributions (SED) and spectra of the early-type galaxies in
RDCS1252, the cluster at $z=1.24$ with the best available multi-wavelength data
set in the cluster sample of early-type galaxies (Demarco et al. 2007).  When comparing with 
the field data we use the spectroscopic sample of early type galaxies in the 
GOODS-S field from the Vanzella et al. (2008) study, with masses determined from the 
multi-wavelength data set offered by the GOODS
survey (Rettura et al. 2006).

\section{Results}

\subsection{The Color-Magnitude Diagram of Cluster vs Field Galaxies}

The color-magnitude diagram ($U-V$ color vs. absolute $V$ magnitudes
in the Vega system) obtained from the model is shown in fig. 1 for
both field (left) and cluster galaxies (right panel) in three redshift
bins. The color code corresponds to the density of galaxies in a given
color and luminosity bin divided by the density of galaxies in the
considered luminosity bin. Note that the cluster color-magnitude
diagram does not correspond to galaxies residing in a single cluster,
but it refers to all galaxies residing in host DM halos more massive
than $10^{14}\,M_{\odot}$.

\begin{center}
\vspace{-0.2cm}
\scalebox{0.65}[0.65]{\rotatebox{0}{\includegraphics{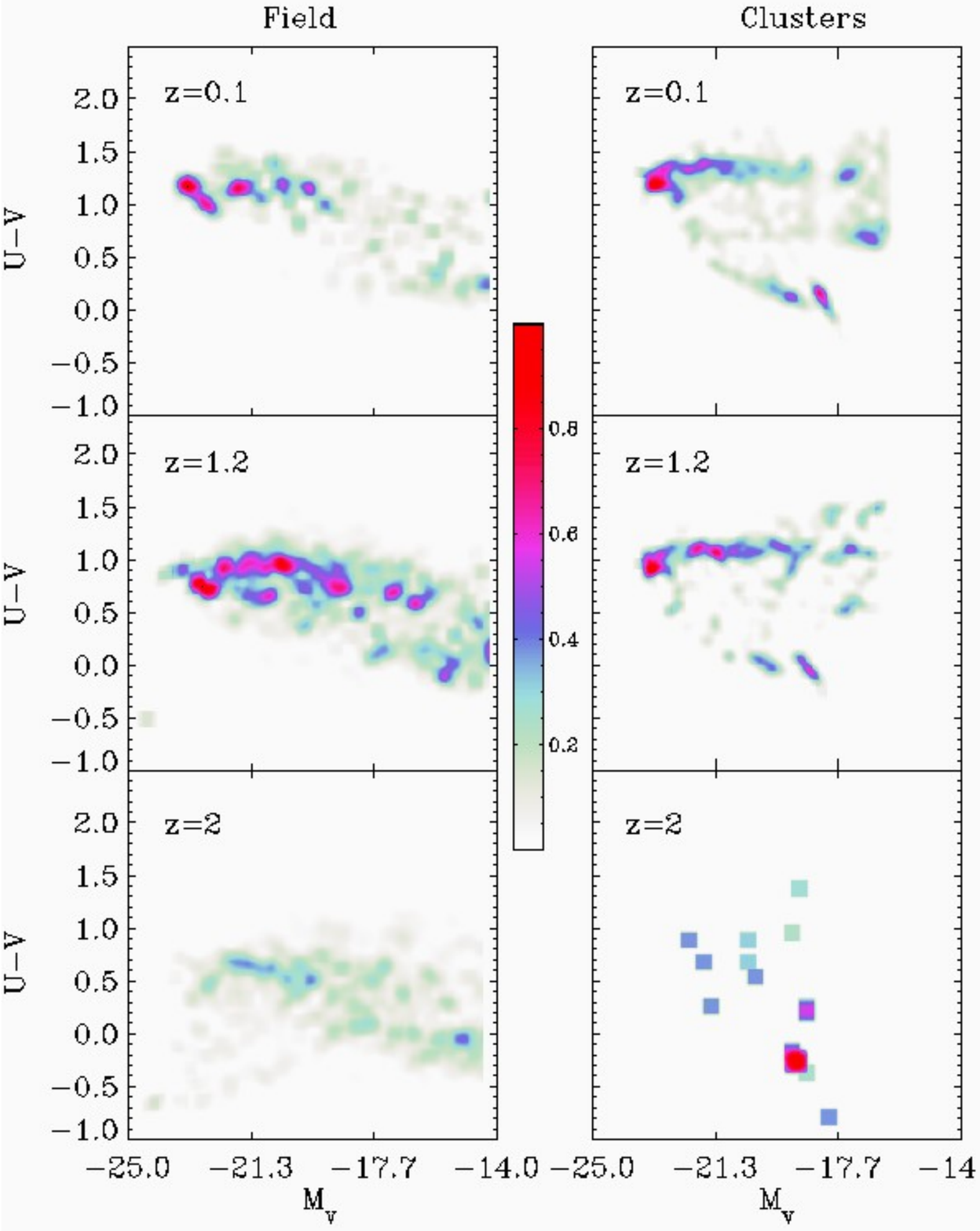}}}
\end{center} {\footnotesize \vspace{-0.6cm }
 Fig. 1. - 
 The color-magnitude ($U-V$ vs. $V$) relation computed from
our model for all galaxies (left panels) and for galaxies hosted in
rich groups or clusters (i.e., in halos with mass $M\geq
10^{14}\,M_{\odot}$, right panels), at $z=0.1$, $z=1.2$, and $z=2$
(from top to bottom). The color code (shown in the central color bar)
represents the number of galaxies in a given color- magnitude
bin normalized to the abundance of galaxies at the considered
magnitude.
\vspace{0.4cm}}

Before discussing in detail the features of the color-magnitude
relation, we first check that the magnitude distribution of galaxies
predicted by the model is consistent with the observations. While at
low redshift the abundance of red galaxies in dense environments (as
well as in the field) has been checked against observations in Menci 
et al. (2005), here we explore it at the highest available redshift in fig. 2,
where we perform a detailed comparison of the $z=1.2$ K-band luminosity
function from our model with the available observations of high-redshift 
clusters (Strazzullo et al. 2006).  
The model field luminosity function is also shown as a dashed line; 
for graphical reasons, its actual normalization (predicted by the model), has been rescaled 
to match the cluster luminosity function at the characteristic 
luminosity $M_{K*}$. 

For the model cluster luminosity function shown in fig. 2, 
we have checked that its normalization is consistent with the observed abundance of bright 
galaxies, both at low and at high redshifts. 
Indeed, the model predicts a cosmic density (per Mpc$^3$) of bright ($M_K<-20$) galaxies, 
located in clusters with mass $M>10^{14.5}M_{\odot}$ at $z=1.2$ (corresponding to
the average X-ray luminosity of the cluster sample adopted in
Strazzullo et al. 2006 using the same conversion adopted by such
authors), which is 3\perc of the overall (field) density of model galaxies
with same brightness at such redshift.  
This ratio between the cluster and the field density predicted by the model 
is in good agreement with the observational value estimated by Strazzullo et al. 
(2006) at the same redshift (4-5 \perc of bright galaxies are found in clusters). 
Thus, a small uncertainty (a logarithmic factor 0.15) in the relative
normalization of the observed and the model luminosity function in
fig. 2 is present. It is interesting to note that the broad agreement
between the model and the observations concerning the fraction of
bright galaxies located in massive clusters persists down to redshift
zero, where the observations yield values of $6-7$ \perc compared to a
predicted value of $\approx 10$ \perc. Finally,  we also show in fig. 2 the  
model luminosity function predicted for the field at the same redshift. Note that 
the model luminosity function shows little dependence on the environment 
(in agreement with observations by Strazzullo et al. 2006), 
although the model yields a somewhat
larger abundance of massive galaxies in clusters compared to the
field. 

\begin{center}
\vspace{0.2cm}
\scalebox{0.5}[0.5]{\rotatebox{0}{\includegraphics{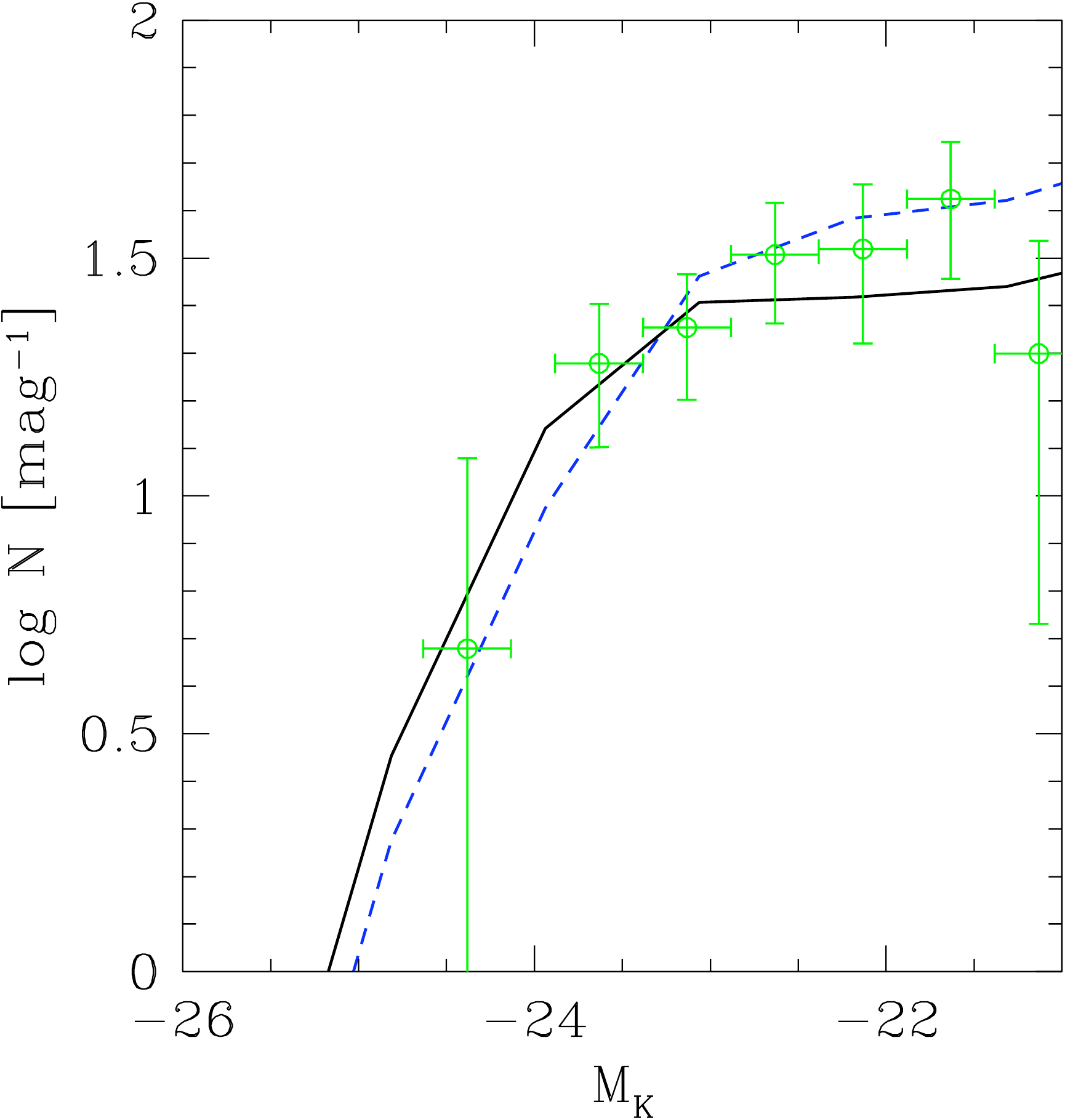}}}
\end{center} {\footnotesize \vspace{0cm }
 Fig. 2. - 
 The K-band luminosity function at $z=1.2$ from our model (solid line) is
compared with data obtained from near-infrared images of three distant X-ray
clusters with an average redshift $z\approx 1.2$ (Strazzullo et al. 2005). 
We have checked the normalization of the model luminosity functions by comparing the 
abundance of bright model galaxies with the observations at both low and high 
redshifts (see text). For comparison, we also show the
luminosity function of model field galaxies at the same redshift $z=1.2$ (dashed line). 
For graphical reasons, its actual normalization (predicted by the model), has been rescaled 
to match the value of the cluster luminosity function at the characteristic 
luminosity $M_{K*}$. 
 \vspace{0.4cm}}

Having checked the abundance of galaxies as a function of their
magnitude against observations, we now examine in detail the
properties of the color-magnitude distribution shown in fig. 1.

Firstly, we note that the overall trend of brighter galaxies to be
redder than fainter ones is effectively modulated by the environment,
so that the fraction of faint galaxies with red colors is larger in
clusters than in the field up to $z\approx 1.5-2$. This reflects the
shift of color distribution of faint galaxies toward red colors in
dense environments already shown for local
galaxies (Menci et al. 2005), and constitutes a first-order consistency-check with
local observations (Baldry et al. 2004) and with the observed trend in
groups up to $z\approx 1.3$ (Gerke et al. 2007).
Here such a check is extended to the richest environments at higher
redshifts and to a wider magnitude range in fig. 3. In particular, the
local fraction of blue galaxies ($f_b$, see caption for the
definition) is shown as a function of absolute magnitude for small
($M/M_{\odot}\leq 10^{13}$) and for large ($M/M_{\odot}\geq 10^{14}$) 
host halo masses, and compared with the SDSS data 
group catalogue (Weinmann et. 2006) in fig. 3a. The decrease of $f_b$ 
with increasing density of the environment, especially marked for faint 
galaxy magnitudes, is not
unexpected in hierarchical theories (see point i) in Sect. 1). It 
traces back to the properties of the primordial density field:
galaxies in groups/clusters are formed from the collapse of clumps
within biased, high density regions of the primordial field, where the
enhanced density allowed for an earlier (and higher) star formation
activity. The gas conversion into stars is thus anticipated and the
rate of galaxy transition from the blue cloud to the RS is accelerated,
leading to a higher fraction of red galaxies at $z\lesssim 2$.

\begin{center}
\vspace{-2.cm}
\scalebox{0.45}[0.45]{\rotatebox{0}{\includegraphics{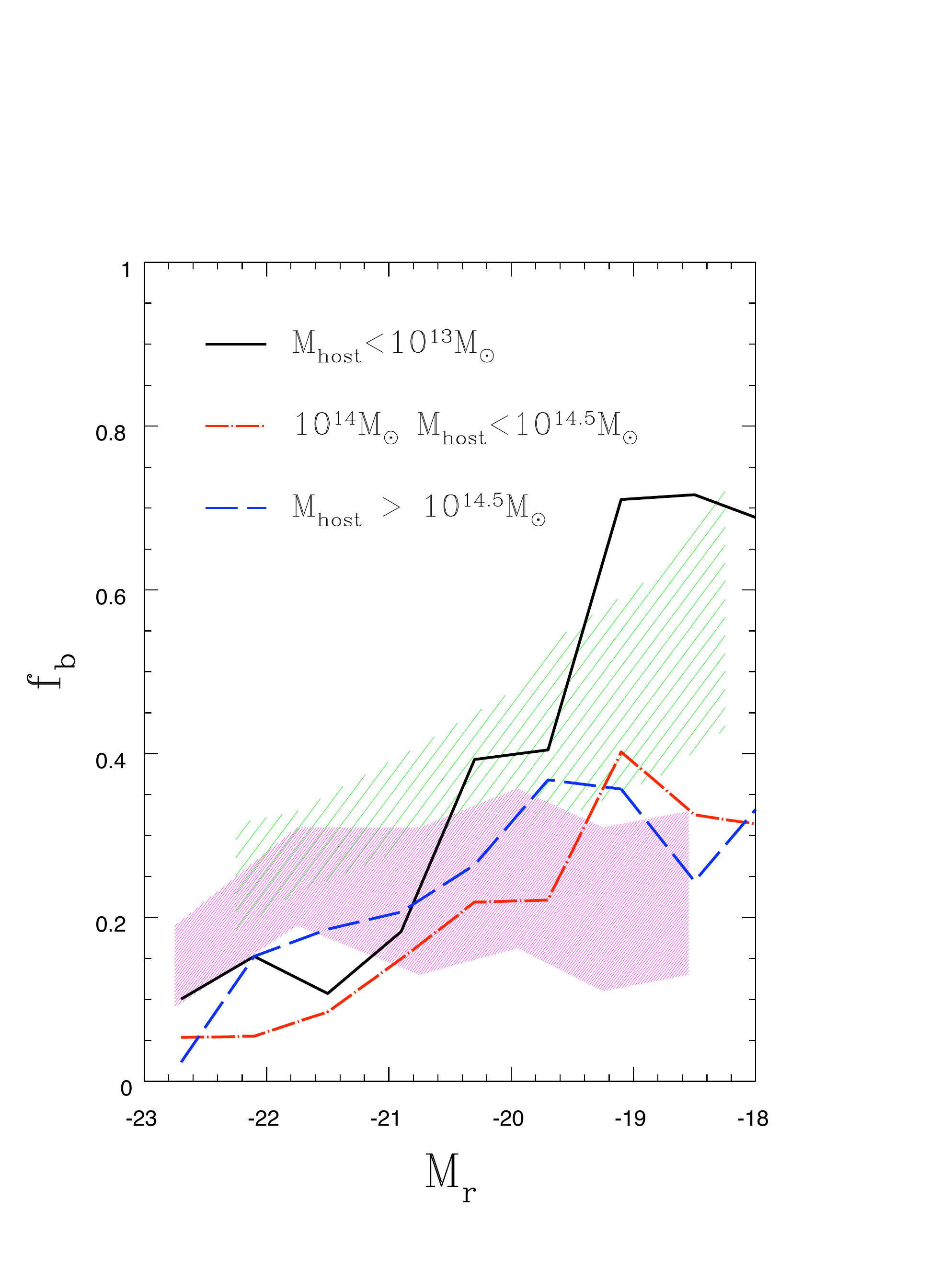}}}
\scalebox{0.45}[0.45]{\rotatebox{0}{\includegraphics{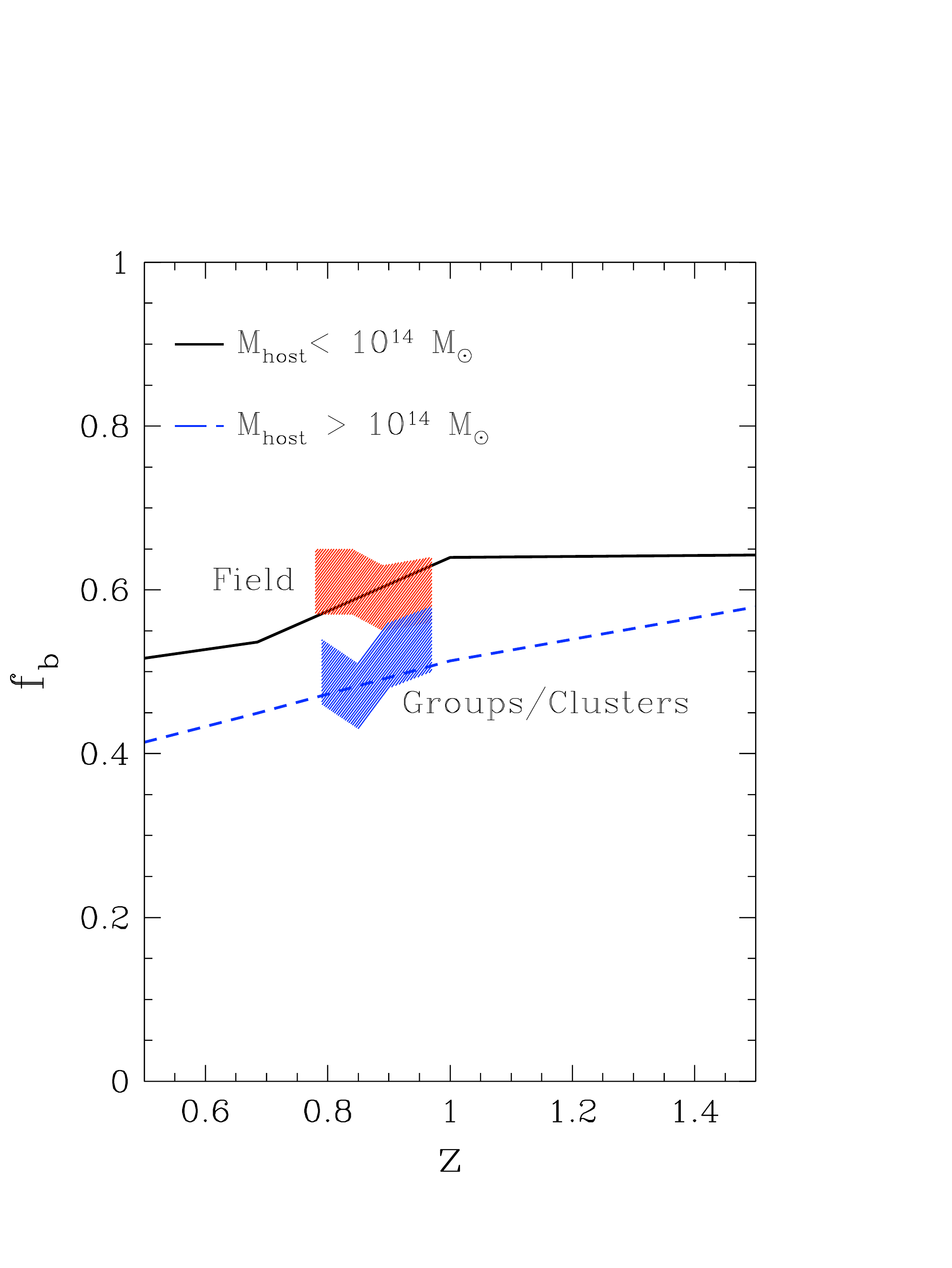}}}
\end{center} {\footnotesize \vspace{-0.5cm }
 Fig. 3. - 
The model blue fraction (lines) is compared with available data for
different environments as a function of the $r$-band absolute magnitude in the
local Universe (left panel) and of redshift (right panel). To perform a proper
comparison, the blue fraction $f_b$ in the two panels has been
computed adopting the same magnitude cut adopted in the papers we
compare with, as detailed below. \newline Left panel: the fraction of
$z=0.1$ galaxies bluer than $(g-r)_{cut}=0.7-0.032(M_r+16.5)$ (where
the SDSS absolute magnitudes k-corrected to $z=0.1$ are adopted as in
Weinmann et al. 2006), is shown as a function of $M_r$ for model
galaxies hosted in haloes with different mass as indicated by the different line types.  
For comparison, we have also included the observational results obtained
from the SDSS group catalogue described in detail in Weinmann et
al. (2006): the upper shaded area corresponds to galaxies in host
haloes with mass $10^{12.5}\,M_{\odot}\leq M\leq 10^{13}\,M_{\odot}$,
while the lower one to galaxies in clusters with
$10^{14}\,M_{\odot}\leq M\leq 10^{14.5}\,M_{\odot}$. \newline Right
panel: to compare with the observations from the DEEP2 survey (Gerke et
al. 2007) the fraction of galaxies bluer than $U-B=-0.032(M_B-5 log
h+21.62)+1.2$ and magnitudes brighter than $M_B-5 log
h=-20.7-1.37(1-z)$ is shown as a function of redshift for model field
galaxies (solid line) and for model galaxies in groups/clusters (with halo
mass $M\geq 10^{14}\,M_{\odot}$, dashed line).  The shaded regions
correspond to the results from the sample I in Gerke et al. (2007)
for field galaxies (upper region) and group galaxies (lower region). 
Making  the group/field division of model galaxies at $10^{13} M_{\odot}$
(most of the groups of the DEEP2 survey have masses in the range 
$10^{13}- 10^{14}\,M_{\odot}$) would yield model results close 
to those plotted in fig. 3b, except for a slight upshift of the dashed line,  
corresponding to an increment 0.025 in $f_b$ approximatively constant with z
 \vspace{0.4cm}}

Secondly, we note from fig.1 that at bright magnitudes, the color
distribution is {\it entirely} dominated by red objects (see also
fig. 3a). The origin of such an effect has been discussed in several
papers (see, e.g., Bower et al. 2006; Menci et al. 2005). 
On the one hand, the downsizing (i.e., the
color/age-luminosity correlation) is a natural outcome of
hierarchical models since star formation within the progenitors of
massive galaxies was enhanced at high redshift due to the biasing effect
discussed above. On the other hand, the probability of grazing
encounters triggering bursts of star formation (see point ii) in
Sect. 1), is larger for massive galaxies (with larger cross section)
at high-redshift (where the densities are larger), thus contributing
to the early conversion of gas into stars for massive group/cluster
member galaxies. However, a substantial fraction ($\approx 40$
\perc, see Menci et al. 2006) of blue, massive galaxies would
persist without the inclusion of AGN feedback in the model. This
effectively suppresses residual star formation in massive halos thus
determining the extremely low fraction of massive (bright) blue
galaxies.

At higher redshift, the overall redshift-dependence of the blue fraction
for bright galaxies (see caption) in clusters and in the field is
shown in fig. 3b.  This shows that the blue fraction predicted in clusters
is larger than in the field up to $z=1.5$. This is not unexpected
since high-redshift clusters (though rare) constitute the most biased
environment for galaxy formation at high-redshift, where the
acceleration of star formation at $z>2$ due to the processes mentioned
above is maximally effective.
Our results concerning the transition of faint galaxies to the red sequence 
in clusters of galaxies are consistent with the findings of other 
semi-analytic models (De Lucia et al. 2007), and extend to higher redshifts
$z\approx 1.2$; our results are also consistent with existing hydrodynamical N-body
simulations (Romeo et al. 2008).   

Thus, the rate of transition of galaxies from the blue cloud to the
red sequence is reasonably well described in the hierarchical
clustering model, so that it can be reliably used to gain insight on
the properties of the RS galaxies on which we focus next.

\subsection{The Properties of the RS in Clusters of Galaxies}

The most striking feature of the color-magnitude diagram
resulting from the model (fig. 1) is the extremely reduced scatter of
the RS in clusters compared to the field, not only at low redshift but
even at $z>1$. When the cluster-to-cluster distribution of the RS
scatter and normalization are examined (see fig. 4), we find that all
clusters are predicted to have a RS color scatter 
$\langle (c-\langle c\rangle)^2 \rangle^{1/2}$ (where $c\equiv U-V$) 
lower than 0.2 mag and that the region enclosing 75 \perc of the model clusters includes values of
the RS scatter $\langle (c-\langle c\rangle)^2 \rangle^{1/2}$ smaller
than 0.1 mag. Such very narrow RS is predicted for all clusters up to
$z=2$.  The high-$z$ values of the scatter and normalization of the RS
resulting from the model are compared in detail with those observed
in the distant cluster sample (sect. 2.2) in fig. 5.

\begin{center}
\vspace{0.2cm}
\scalebox{0.75}[0.75]{\rotatebox{90}{\includegraphics{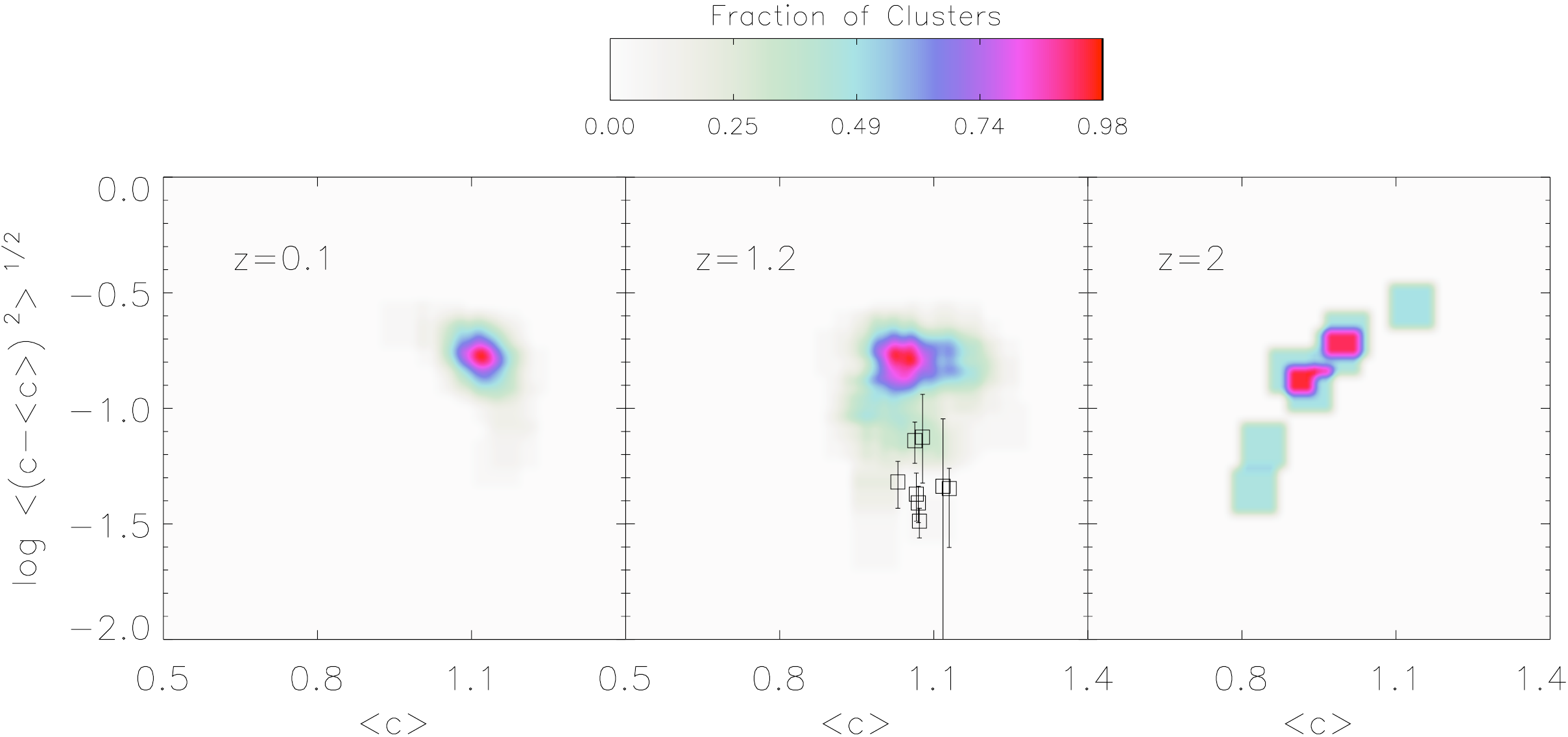}}}
\end{center} {\footnotesize \vspace{0.cm }
 Fig. 4. - 
 Normalization and scatter of
of the RS among the modeled cluster population. The RS is defined as
the region of the color-magnitude plane (fig. 1) with $c\equiv
U-V>0.8$, and for this we compute the average color $\langle
c\rangle$ and the dispersion $\langle (c-\langle c\rangle)^2
\rangle^{1/2}$ around the average. The applied color cut is
consistent with the minimum in the color distribution which
separates the two peaks corresponding to the blue and the red
populations (see fig. 1 and the color distribution in the AB
magnitude system computed in Menci et al. 2005, 2006).
The color code in the figure corresponds to the fraction of model clusters with given 
RS scatter $\langle c\rangle$ and normalization $\langle (c-\langle c\rangle)^2\rangle^{1/2}$.
For comparison, we also show in the same plane the points that correspond
to 8 observed clusters in the redshift range $0.8\leq z\leq 1.3$ from the papers 
cited in Sect. 3.
 \vspace{0.4cm}}

Note that an exact comparison between observed and model RS scatter  
constitutes a delicate point, since the observational morphological selection 
and the outlier rejection criteria adopted in the data analysis may affect the comparison with the model, 
which includes exact cluster members with any morphological types. 
In the original analysis of observed cluster data performed in the papers cited in Sect. 3, 
only early-type galaxies were included: the corresponding observational results are represented by the square points (with the errors given by the smaller error bars) in fig. 5. 
For the sake of a more accurate comparison to the model, we have recalculated the 
RS scatters for the same observed sample using a procedure as close as possible to the one adopted 
for model galaxies. In particular: 1) we included 
all red sequence (rest--frame $U-V> 0.8$) galaxies, irrespective of their morphological type, as in the model; 
2) recomputed isophotal colors as from Sextractor (Bertin \& Arnouts 1996); 
3) performed a 3--$\sigma$ cut selection and made use of the available spectroscopic redshifts where 
available to reject outliers. 
The observational RS scatters are computed as $\langle (c-\langle c\rangle)^2 \rangle^{1/2}$ as done for model galaxies. 
The results from this new analysis of cluster data are to be considered as upper limits to the 
actual RS scatters due to the contamination from outliers. The larger possible values for the 
observational RS scatters (including the uncertainties) resulting from our new analysis are shown as 
downward arrows in Fig. 5. 
The exact observational RS scatters to be compared with the model would lie between these upper limits 
and the squared points, and would have to include spectroscopically confirmed members for all morphological 
types (data that we do  not have at present).

\begin{center}
\vspace{-2.cm}
\scalebox{0.5}[0.5]{\rotatebox{0}{\includegraphics{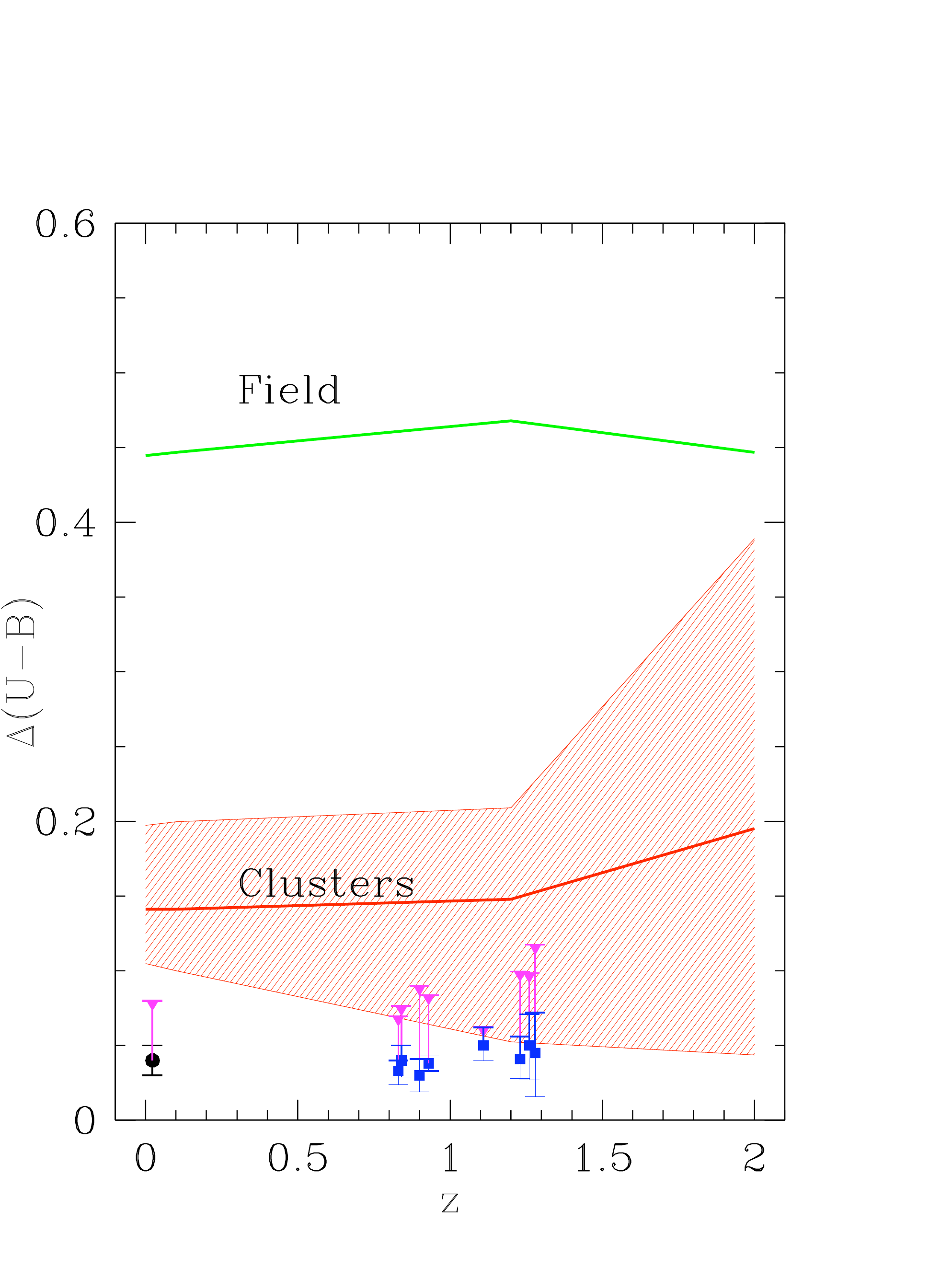}}}
\end{center} {\footnotesize \vspace{-0.6cm }
 Fig. 5. - 
 Comparison between predicted and observed scatters of the RS
in high-redshift clusters. The lower shaded area represents the
region enclosing 75 \perc of the clusters in our Monte Carlo
computation; the observational data points (squares) are taken from
Blackeslee et al. (2003, 2006), Homeier et al.  (2006), Mei et
al. (2006a,b). The low redshift point corresponds to the  
RS $U-V$ scatter of the Coma cluster measured by Bower, Kodama, Terlevich (1998). 
The downward arrows above all data points indicate the upper limits   
on the observed RS scatter obtained from our new data analysis, 
which includes galaxies of any morphological type 
(see text). For comparison, we also show our estimate of the
scatter of the RS for model field galaxies (upper solid line).
 \vspace{0.3cm}}

Although appreciably smaller than the corresponding field value, the model 
RS scatter $\langle (c-\langle c\rangle)^2 \rangle^{1/2}$ for 
clusters is generally slightly larger than the observed
values at high redshifts, which lie in the lower envelope of the
region enclosing 75 \perc of the model clusters (the light-blue
region in fig. 4 and the shaded lower area in fig. 5). It is also
interesting to note that the model predicts a RS of high-z
clusters with no universal average color, but rather with cluster-to-cluster variance  
of about 0.2 mag in $\langle c \rangle$. This is expected
since at high redshift clusters are still forming, so that the
distribution of average colors reflects the different formation
paths of these structures which are just emerging from the density
field. Such a prediction is in agreement with the HST/ACS
observations as is shown in the central panel of fig. 4, where the
points corresponding to observed clusters show RS colors differing
by $\approx 0.2$ mag. Note that the average value of the RS 
scatter in the model decreases with redshift from z=2 to z=0 (see the red continuous line in fig. 5). 
The dispersion around such an average scatter decreases with redshift (see shaded area in fig. 5 and fig. 4) 
since different merging histories of sub-clusters formed at high redshift with different masses 
are included into a unique rich cluster at low redshift. 

Thus, the model seems to capture the basic processes responsible for the 
narrow scatter of the RS of clusters, although the
observations still lie on the envelope of the distribution of
simulated clusters, which have a mean scatter at least twice
as large as the observed one. The origin of this general behavior
must be traced back to the fact that RS galaxies in high-redshift
clusters are those formed in the most biased environmen existing at
those redshifts, so two possibilities arise to explan 
such an effect. 1) The peak of the {\it star formation history} of
RS galaxies in clusters is shifted to earlier cosmic times compared
to RS galaxies in the field, thus compressing the star formation activity 
of the progenitors of RS cluster galaxies in a narrow time range. 
In this case, the dense environment would not only
accelerate the transition from the blue cloud to the red sequence
(thus depressing $f_b$ in clusters) but also affect the ages of
galaxies which have already moved to the RS. 2) The effect is due to
an intrinsic lower scatter in the {\it mass assembly history} of RS
galaxies in clusters with respect to the field.  This means that the
number of possible paths along the merger trees leading to RS
galaxies is smaller for galaxies in clusters compared to the field.
We investigate both possibilities below.

\subsection{The Mass Growth History and The Age Distribution of
Cluster vs. Field Galaxies}

In fig.~6, we show the model predictions for the mass growth histories
of field and cluster galaxies for galaxies belonging to the RS
(i.e., with color $U-V\geq 0.8$) with stellar masses in the range
$5\times 10^{10}\leq M_*/M_{\odot} \leq 5\times 10^{11}$.  The color
code refers to the fraction of galaxies which (at the given cosmic
time shown in the x-axis) has attained the fraction of final stellar
mass shown in the y-axis. The observational results (square data
points with error bars) are from Gobat et al. (2008) who used two
samples of spectroscopic early-type galaxies in the same stellar mass
range drawn from the GOODS field (Vanzella et al. 2008) and the
massive cluster RDCS1252 at $z=1.24$ (Demarco et al. 2007), and
derived the star formation histories by fitting both their spectral 
energy distributions and spectra with a large grid of
Bruzual-Charlot (2003) models.

Inspection of fig. 6 shows that in the hierarchical model the {\it
  average} stellar mass growth of cluster galaxies does not differ
substantially from the field {\it when only red sequence galaxies
  are considered}; it is only the {\it fraction} of galaxies belonging
to the RS which strongly depends on the environment, as discussed in
the previous section. Comparison with the data shows a global
consistency between expected and observed growth histories, although
observations suggest slightly earlier stellar mass formation in
cluster galaxies compared to the model. We should stress, however,
that while the model mass growth histories are computed
self-consistently from the star formation histories resulting from
Monte Carlo merging tree simulations, the data points have been
obtained by fitting star formation histories based on
Bruzual-Charlot models with adjustable parameters affected by some
degeneracies.

\begin{center}
\vspace{0.2cm}
\scalebox{0.8}[0.8]{\rotatebox{90}{\includegraphics{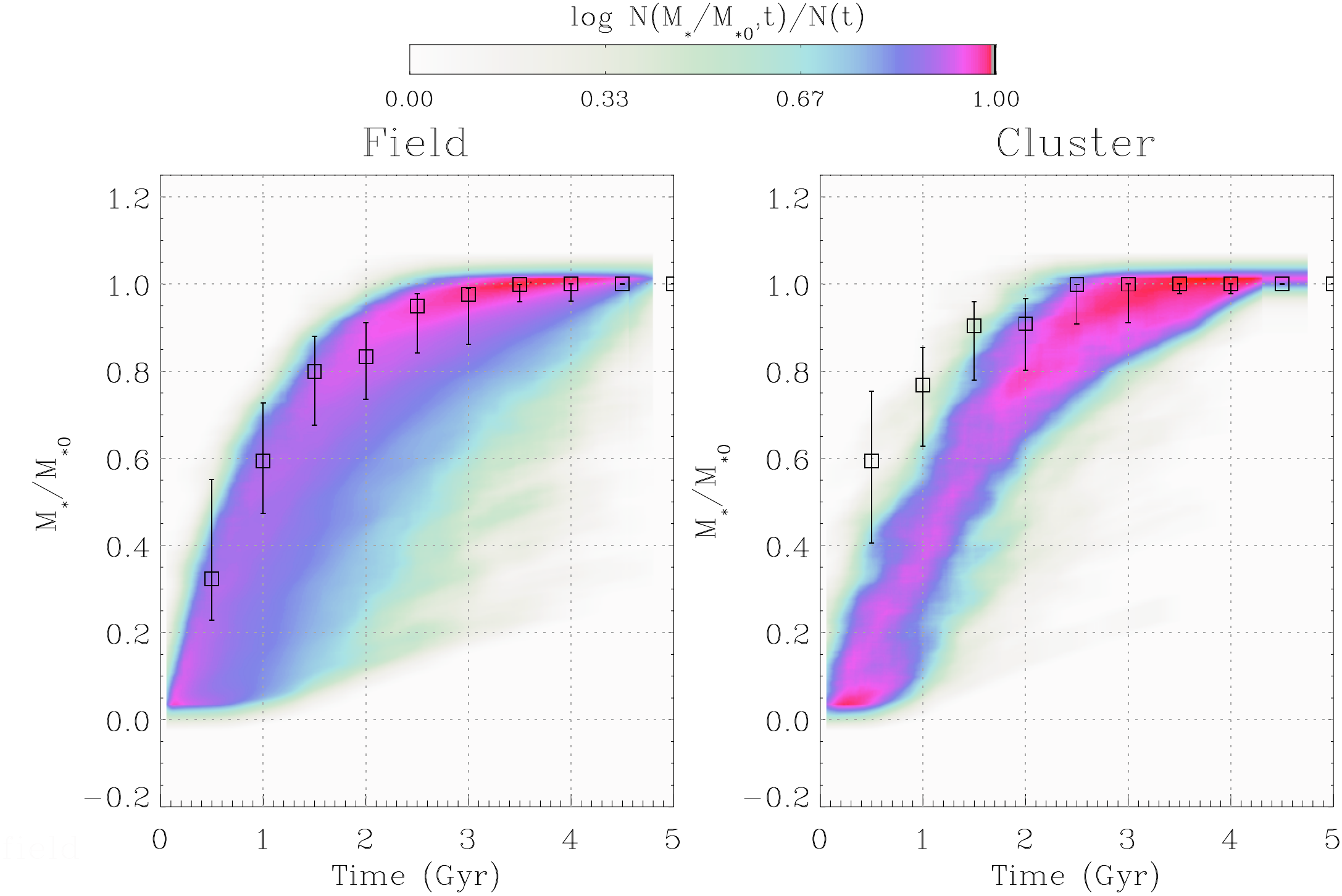}}}
\end{center} {\footnotesize \vspace{0.cm }
 Fig. 6. - 
 The stellar mass growth history of galaxies at $z=1.2$ with stellar masses in
the range $5\times 10^{10}\leq M_*/M_{\odot} \leq 5\times
10^{11}$. The mass fraction $M_*/M_{*0}$ (y-axis) formed at the
cosmic time t (x-axis) is computed as the stellar mass contained in
all progenitors of each considered galaxy at $z=1.2$ divided by its
final (at $z=1.2$) stellar mass. The color code represents the
number of galaxies with given $M_*/M_{*0}$ at a cosmic time $t$
normalized to the total number of galaxies at that time. The data
points, from Gobat et al. (2007), are derived from
spectrophotometric observations of early-type galaxies in the field
and in a rich cluster at $z=1.24$, with stellar masses in the same mass
range.
 \vspace{0.4cm}}

Fig. 6 also shows that a major difference between the model cluster and field
galaxies is in the {\it scatter} of the mass growth histories, which
is much smaller for cluster galaxies compared to the field. For
example, while there is a substantial fraction of field galaxies which
has assembled only 60 \perc of their stellar mass after 2 Gyr, almost
all cluster galaxies have assembled more than 75 \perc of their
stellar mass at the same cosmic time. Thus, the small scatter observed
in the cluster RS seems to be originated from a smaller dispersion in
the mass growth histories of galaxies rather than from an earlier
overall stellar mass growth. This is confirmed by the analysis of the
distribution of stellar ages shown in fig. 7. We define star formation
weighted ages as
\begin{equation}
\tau(t)={\int_0^t\,dt'\,(t-t')\dot m_*(t')\over \int_0^t\,dt'\dot m_*(t')}\,\,\,,
\end{equation}
where $t$ is the cosmic time at which we compute the average age of
the stellar populations, and $t-t'$ is the age of the populations
formed at time $t'$ at a rate $\dot m_*(t')$.
This definition has the advantage of taking into account the effective fraction of
stellar mass contributed by each stellar population included in the average. 
Thus, the ages of the stellar populations contributing only a negligible fraction of the final
stellar mass do not affect appreciably the average age defined by (1).

The predicted distribution of such quantity among the RS galaxy population in
field and clusters at $z=1.2$ is shown in the figure for the whole
range of galaxy masses (bottom panel) and for the stellar mass range
$5\times 10^{10}\leq M_*/M_{\odot} \leq 5\times 10^{11}$ (top panel)
used in fig. 6.

The bottom panels of Fig. 7 show an evident downsizing in both the
field and cluster distributions, in the form of increasing average
stellar ages with increasing galaxy stellar mass. This is consistent with 
a parallel analysis of observations by Rettura et al. (2008) of the same field and
cluster samples used by Gobat et al. (2008). The top panels also show a good
agreement between model predictions for SF weighted ages and those
derived from the spectrophotometric data (Gobat et al. 2008).  In
particular, the peak values of the stellar ages in the field and in
clusters do not differ substantially in the considered mass range (a
confirmation of our results on the stellar mass growth). However,
while the cluster galaxies have a distribution neatly peaked around
the median value, the age distribution of field galaxies shows a
larger scatter with a tail toward shorter ages; the fraction of
galaxies with ages shorter than 2 Gyr is negligible in clusters but is
around 0.25 in the field.

\begin{center}
\vspace{0.2cm}
\scalebox{0.7}[0.7]{\rotatebox{90}{\includegraphics{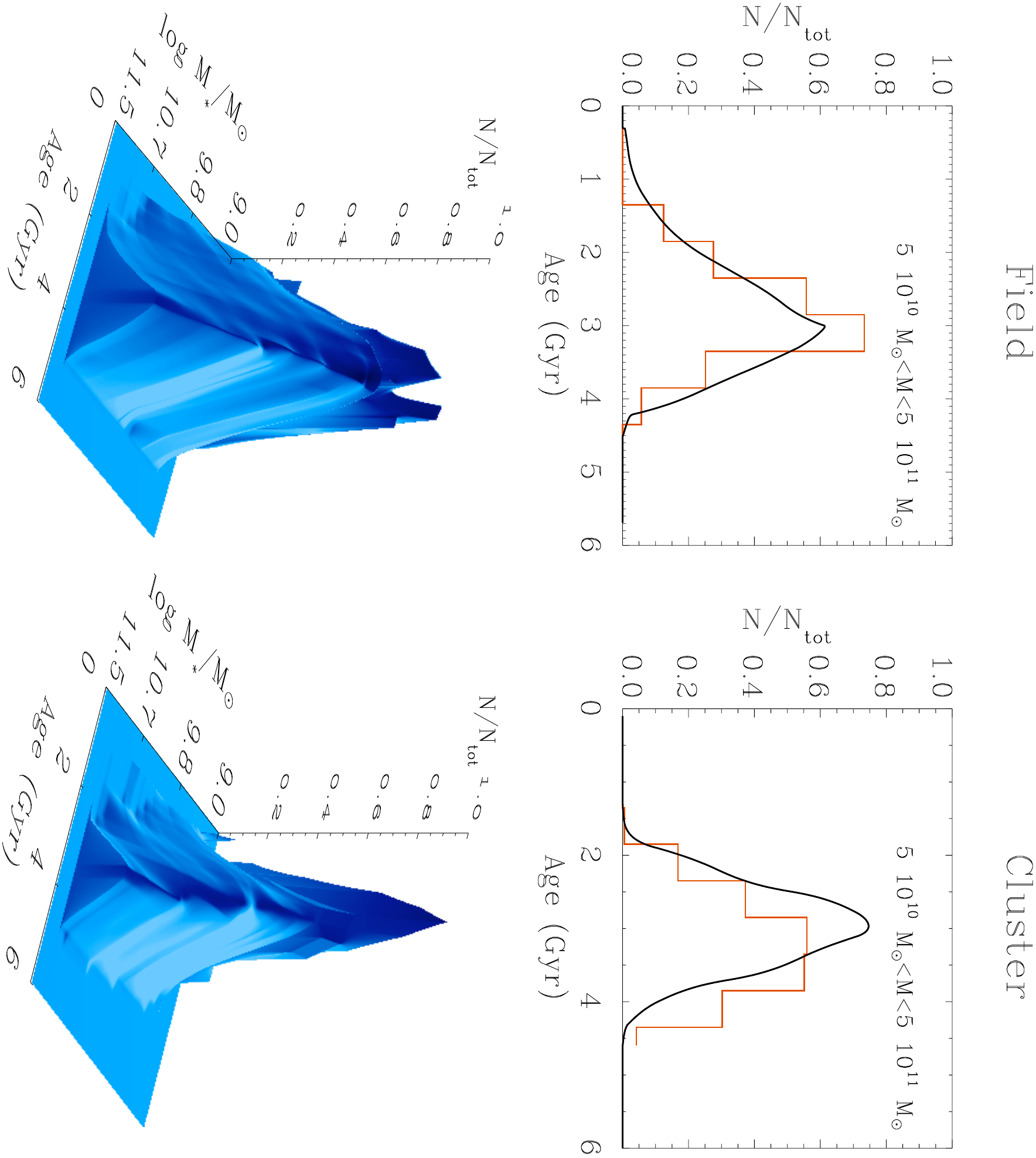}}}
\end{center} {\footnotesize \vspace{0.cm }
 Fig. 7. - 
The distribution of stellar ages (see definition in the text) as a
function of final stellar mass for galaxies in the field (left) and
cluster (right-hand) at $z=1.2$. \newline Top Panels: the
distribution of ages (solid lines) for galaxies in the mass range
$5\times 10^{10}\leq M_*/M_{\odot} \leq 5\times 10^{11}$ at $z=1.2$
in the field (left-hand) and in clusters (right-hand). The histograms 
represent the observed distributions of stellar ages derived, 
using eq. (1), from fitting with Bruzual-Charlot models the same 
spectrophotometric data as in fig. 6 (Gobat et al. 2007).
\newline
Bottom Panels: The distribution of stellar ages (see definition in the text) 
as a function of final stellar mass for galaxies in the field (left) and clusters 
(right) at $z=1.2$. 
 \vspace{0.4cm}}

Thus, both observational and model results concur in indicating that 
the smaller scatter of the RS in rich clusters compared to the field at high-$z$ 
cannot be accounted for by the difference
in stellar mass assembly between field and cluster RS galaxies; however, such a
(tiny) difference (not reproduced by model, see right panel in fig. 6)
may be responsible for the residual discrepancy (a factor
$\approx 1.5$) between the observed and the model RS, shown in figs. 4 and 5.
We shall come back to this point in the Conclusions.

\subsection{Discussion: the Origin of Reduced Scatter in the Stellar
Mass Assembly of Cluster Galaxies}

To understand the origin of the reduced scatter of the ages, color,
and mass growth histories of clusters galaxies compared to the field
we show in fig. 8 (top panel) the model resuts concerning the 
ratio between the star formation 
contributed by minor progenitors (i.e., all progenitor DM haloes except the
main one) to that contributed by the main progenitor, as a function of
cosmic time and final (at $z=1.2$) galaxy stellar mass. Clearly, minor
progenitors of cluster galaxies have contributed significantly to the
stellar mass growth only at very early epochs, when they had all
comparable masses. At times $t\gtrsim 1$ Gyr the contribution from
minor progenitors is negligible, and the bulk of the stellar mass is
formed in the main progenitors. Conversely, for field galaxies the
star formation in minor progenitors remains appreciable even at later
cosmic times, and the scatter generated by the different merging
histories effectively results into a final dispersion of galaxy color,
ages, and stellar mass growth histories.

\begin{center}
\vspace{0.1cm}
\scalebox{0.65}[0.65]{\rotatebox{0}{\includegraphics{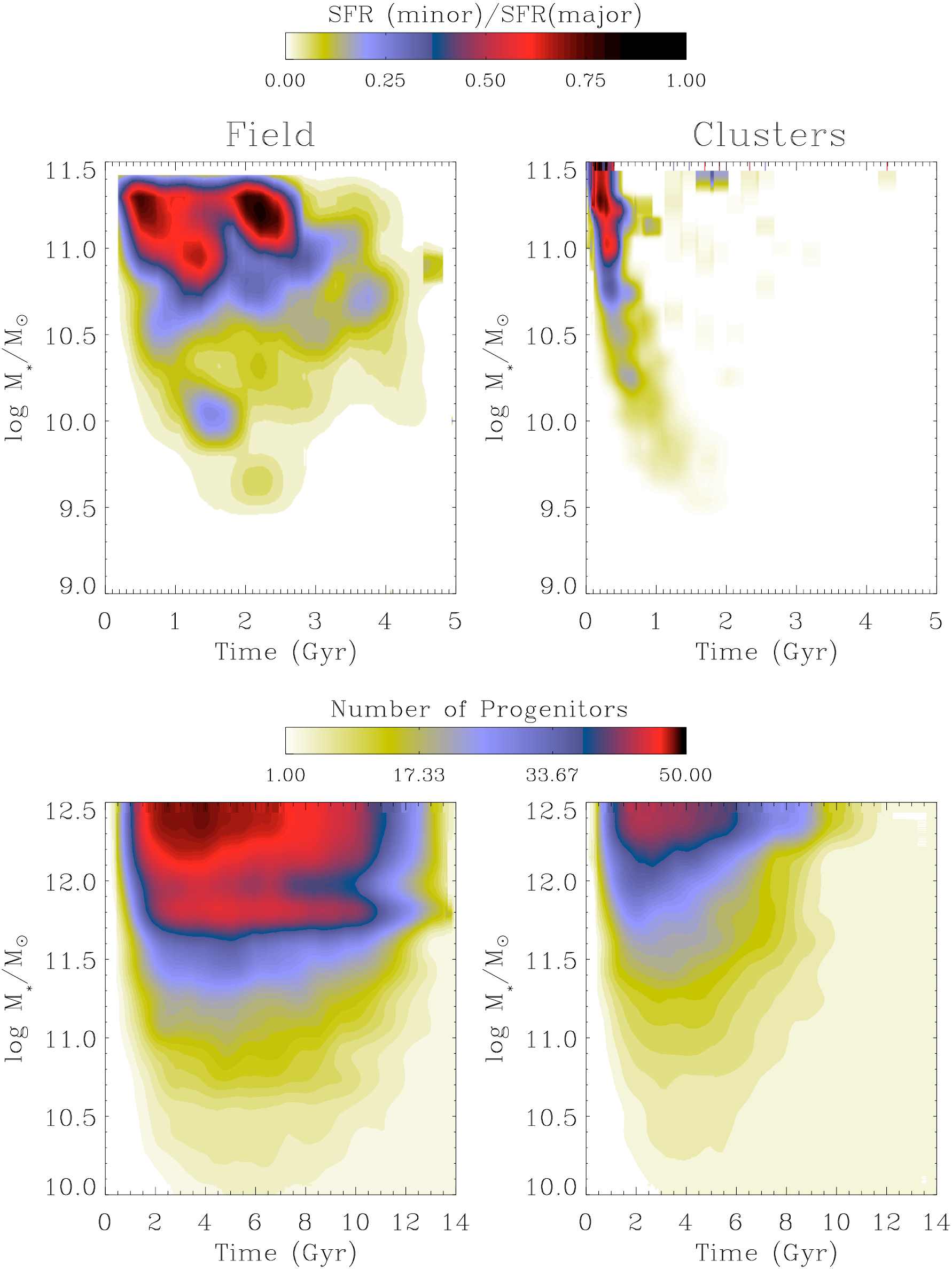}}}
\end{center} {\footnotesize \vspace{0.cm }
 Fig. 8. - 
 Top panels: The ratio between the star formation rate contributed by minor
progenitors (see text for definition) and the main one (color code) is plotted
as a function of the cosmic time (x-axis) for different values of the final (at
$z=1.2$) stellar mass (y-axis). Left-hand column refers to field galaxies,
right-hand column to cluster members.\newline Bottom panels: The number of
progenitors with stellar mass larger than $5\,10^9\,M_{\odot}$ is represented by
the colored contours as a function of cosmic time $t$, from $t=0$ to the present
for different values of the final (at $z=0$) stellar mass (y-axis).
 \vspace{0.4cm}}

In other words, the model predicts that, for field galaxies, the formation of stellar
populations finally included in a given galaxy is largely distributed
among all its progenitors, while the majority of galaxies in clusters
have been assembled earlier on into a single main progenitor, which
hosts the bulk of the subsequent star formation.

Thus, although the overall (contributed by all progenitors) median 
star mass growth of field and cluster galaxies is similar (see
figs. 6, 7) the scatter is considerably suppressed for the latter,
since the variance introduced by the different merging histories is
considerably reduced.

For completeness, we also show in the bottom panel of fig. 8 the cosmological
evolution over the whole cosmic time of the predicted number of progenitors with
stellar mass larger than $5\,10^9\,M_{\odot}$. Note how, for field
galaxies, this number remains large even at low redshift, indicating a
delayed assembly of progenitors into a main object compared to cluster
galaxies. Note also that a small number of massive progenitors is
accreted also by cluster galaxies. Since the top-right panel shows
that their contribution to the build up of the stellar mass is
negligible when compared to the main progenitor, this implies that
most of the minor clumps accreted by cluster galaxies are gas poor, a
condition due to the gas stripping caused by the surrounding
cluster medium.

Thus, the different scatter of the RS between clusters and field 
results from the different balance between dry merging 
and star formation history of progenitors. That such a balance 
constitutes a key point in determining the properties of RS galaxies has been 
recently pointed out by Faber et al. (2007), who proposed 
that the building up of the RS results from a mixture of early star formation
followed by quenching and dry merging, and noticed that 
the prevalence of the latter would result in a reduced scatter of the RS. 
Here we argue that the balance between the 
two processes depends on the environment, with dry merging of progenitors with quenched 
star formation prevailing in biased environments (as results from the compution 
shown above). This leads in clusters to a relevant reduction not only of the 
RS scatter, but also of the scatter of the ages of the 
stellar populations of RS galaxies, a marker of the 
effect of dry merging already noted by Faber et al. (2007). 

\section{Conclusions}

In this paper we have compared the results from a state-of-the-art
semi-analytic model of galaxy formation to spectrophotometric
observations of the most distant galaxy clusters observed in the range
$0.8\lesssim z\lesssim 1.3$, to investigate the properties of the red
sequence (RS) galaxies in high-redshift clusters and to compare them
with those of the field in the same stellar mass range.

A long standing prediction of cosmological galaxy formation models
(see Kauffmann 1996; Diaferio et al. 2001 ) is that the {\it average}
stellar populations of galaxies in clusters are older, since the star
formation processes are accelerated in structures originated from 
overdense regions of the primordial density field. Recent
observational progress (Baldry et al. 2004) has revealed in
detail such an average reddening of galaxies for increasing
environmental density, due to an increased population of the red
branch of the bimodal color distribution. Thus, the redder {\it
  average} colors of the cluster galaxy populations we observe today
are not caused by a global shift of the ages of the galaxy
populations toward lower values, but instead by a different balance
between the populations of the two branches of the color distribution.

In particular our results can be summarized as follows

\begin{itemize}

\item The speed up of the transition of galaxies from the blue cloud
  to the red sequence which takes place in dense environments is
  qualitatively reproduced by the hierarchical model for a wide range of
  galaxy masses (being faster for more massive galaxies) and cosmic
  times. In the case of distant galaxy clusters, the most extreme
  environments which provide the most challenging test for hierarchical
  models, such a fast transition in the model is able to reproduce a
  well defined red sequence in clusters at $z\approx 1.3$, which is
  narrower than in the field, as observed. The fraction of galaxies
  which belong to the RS is larger in clusters than in the field from
  $z=0$ to $z=1.5$, and stays nearly constant in such an interval.

\item In agreement with observations, we find that at $z\approx 1.2$
  the RS is not universal, i.e. is subject to a cluster-to-cluster
  variance over the cluster population. The predicted color
  normalization is close to the observed average value and is
  distributed in a narrow range $\langle c\rangle \approx 0.9-1.1$,
  when different clusters are considered. Observations (e.g. Mei et
  al. 2007) suggest a cluster-to-cluster variation of $\sim\! 0.05$ mag,
  a measurement which is limited by sample size and uncertainties in
  the K-correction and passive luminosity evolution in the redshift
  interval $0.8\leq z\leq 1.3$.

\item The scatter of the model RS of cluster galaxies is found to be
 much tighter than
  that of the field by a factor ranging from 5 to 10. However, on
  average, they are still larger than the observed tight scatters
  ($\approx 0.05-0.08$ mag) which are consistent only with the lower
  envelope of model distribution enclosing 75 \perc of the
  clusters selected in the Monte Carlo simulation.

\item When {\it only} the RS galaxies are considered, the mass growth
  histories of field and cluster galaxies at $z\approx1.2$ are
  similar. This is consistent with the mass growth derived from
  observations through SED and spectral fitting (Gobat et al. 2008) of
  massive early-type galaxies with adjustable star formation
  laws. Both observations and the model converge in indicating that at a 
  cosmic time $t\approx 2.5$ Gyr more than 90 \perc of the stellar mass
  finally assembled in the RS galaxies at $z=1.2$ has already been
  formed in both field and clusters. The model also suggests that at
  cosmic times $t\approx 1$ Gyr
  the stellar mass fraction was about 50 \perc in both field and
  clusters, whereas spectral-synthesis modeling of the
  spectrophotometric data suggests a faster time scale in assembling
  stellar mass in clusters. Even though our model predicts the scatter of
  the growth histories to be larger in the field than in clusters, a
  slightly faster assembly time scale seems to be needed to account for
  the very tight observed RSs.

\item The corresponding predicted ages of the stellar populations in
  RS galaxies at $z\approx 1.2$ reflect the above behavior, with
  distributions peaking at values $\tau\approx 3.5$ Gyr in both field
  and clusters. These are consistent with the
  values derived from observations through SED+spectral fitting,
  although the latter indicate a slightly larger (by $\lesssim 0.5$ Gyr)
  age for cluster RS galaxies as compared with the model (Gobat et al. 2008; 
  Rettura et al. 2008).  The model
  prediction of a larger dispersion of ages for field RS galaxies
  compared to clusters is in good agreement with observations. Indeed,
  a negligible fraction of RS galaxies are predicted and observed to
  have ages shorter than 2 Gyr, while an appreciable - though minor-
  fraction ($\approx 0.2$) of field RS galaxies have ages shorter than
  that.

\item The reduced scatter in color, ages, stellar mass growth, of
  cluster RS galaxies at $z\approx 1.2$ compared to field RS galaxies
  is originated from intrinsic properties of the merging histories of
  cluster galaxies. These are characterized by an earlier (compared to
  the field) assembly of galaxy progenitors into a main progenitor at
  times $t\lesssim 1.5$ Gyr. After such a time, the bulk of the stellar 
  mass formation of cluster RS galaxies takes place in their main
  progenitor; thus the scatter in mass growth histories, ages and
  colors is considerably suppressed, since the variance
  introduced by the different merging histories is considerably
  reduced compared to the field, where the star formation is
  distributed among a larger number of progenitors.

\end{itemize}

The picture emerging from the above results is the following. For
increasing cosmic time, the exhaustion of cold gas in galaxies and the
subsequent decrease of the star formation activity drives the galaxies
to flow from the ''blue cloud'' region of the color-magnitude space to
the RS; the more massive the galaxies (and the earlier their gas
conversion into stars), the earlier is the epoch of drift to the RS.

In dense environments, and in particular for high redshift clusters, 
the drift of faint galaxies
from the blue cloud to the RS is accelerated by various processes (namely,
processes i)-iv) recalled in the introduction) so that also fainter
galaxies rapidly drift to the RS. As a result, the blue fraction in
clusters is significantly smaller than in the field, and at
high-redshift a well defined, narrow RS is already in place.

When only the galaxies that have drifted to the RS are considered,
their {\it average star formation history} does not depend strongly on
the environment and is not responsible for the reduced scatter of the RS
in clusters compared to the field; the effect of the environment is to
accelerate the drift to the RS (thus enhancing the relative population of
the red branch compared to the blue branch in the color distribution)
rather than changing the properties of the RS itself. This is
supported by the detailed measurements of the color distribution of
galaxies as a function of the environment in Baldry et al. (2004).

On the other hand, the {\it mass assembly history} of RS galaxies in
clusters is considerably different from that occurring in the field,
and has a significant impact on the scatter of the RS.  Since cluster
galaxies are formed from clumps that collapsed within high-density 
regions of the primordial density field, their assembly into a
dominant progenitor is faster compared to the field, the formation of the 
bulk of the stellar mass of cluster RS galaxies takes place in their main
progenitor, and the variance introduced to the star formation history
by the different merging histories is considerably suppressed. The
more biased the galaxy environment, the larger the above effect,
so that for RS galaxies in high-redshift clusters the dispersion of
mass growth histories, ages and colors is appreciably reduced compared
to the field.

Despite the overall good agreement between our models and
observations on the color, luminosity and mass distributions of
galaxies in high and low density environments, our analysis has
shown that even faster mass assembly or shorter star formation time
scales are needed to reproduce the surprisingly tight RS in the
massive clusters at $z\approx 1.2$ (see fig. 5). This is
extremely difficult to achieve in the model, since the dynamical history of DM
haloes does not constitute a tunable component for a 
given initial power spectrum. Indeed, we have also verified that
changing the normalization of the power spectrum, while affecting
the abundance of clusters at a given redshift, is not effective in
further narrowing down the scatter of the color-magnitude relation. Thus,
we propose that the acceleration of mass assembly in cluster galaxies
(making the bulk of the effect) must be complemented with a further
acceleration of star formation in RS cluster galaxies. This is
indicated also by the slightly mismatch between model and
observations in the stellar mass growth history (fig. 6) and average
ages (fig. 7), when the RS cluster galaxies are considered. The
modeling of such an additional enhancement of early star formation
in the progenitors of cluster RS galaxies will probably require a
finer treatment of the environmental dependence of the
interaction-driven starbursts.
An additional issue is the slope of the cluster RS obtained from the model. 
Current observations find no significant evolution with redshift of the RS slope 
which remains close to $| \delta (U-B)/\delta B| \approx 0.04$ 
(Mei et al. 2008, submitted), whereas the model predicts a (non-evolving) flat slope. 
This can be due to both an over-quenching 
of star formation at low masses, but also to a poor treatment of chemical evolution 
(e.g., the contribution of SNI is not properly accounted for in semi-analytic models) which is known  to affect the RS slope through metallicity effects (Kodama \& Arimoto 1997).

Finally, we mention two additional consequences of our interpretation of the
model results, both consistent with preliminary indications from observations.
First, at redshift $z\gtrsim 1.5$ our model predicts the scatter of the
cluster RS to extend to a wider range of values
compared to the low-redshift interval $z\lesssim 1.5$ (see fig. 4) and to
easily reach values larger than 0.2 (see fig. 5), consistent with preliminary
results by Gobat et al. (2008). Second, the model predicts the
speed-up of star formation in dense environments to be larger for small mass
galaxies (with $M\lesssim 10^{10.5}\,M_{\odot}$) (as inferred, e.g., by fig. 3)
compared to massive RS galaxies (which indeed are predicted to have similar
mass assembly histories in the field and in clusters, see fig.6). This is
also qualitatively consistent with preliminary results by Gobat et. a. (2008),
which indicate a lower average age of stellar populations in clusters
only for galaxies with $M\lesssim 10^{11.2}\,M_{\odot}$.
We shall address in detail both points in a next paper.

\acknowledgments We thank the referee for helpful comments which contributed to improve the manuscript. We acknowledge grants from INAF and the ESO visitor programme. The National Radio Astronomy Observatory is a facility of the National
Science Foundation operated under cooperative agreement by Associated
Universities, Inc.


\bigskip\noindent

\begin{references}

\reference{} Baldry, I.K., Glazebrook, K., Brinkmann, J., Zeljko, I., Lupton, R.H., Nichol, R.C. and Szalay, A.S. 2004, ApJ, 600, 681

\reference{} Baldry, I.K., Balogh, M.L., Bower, R.G., Glazebrook, K., Nichol, R.C., Bamford, S.P., Budavari, T. 2006, MNRAS, 335, 441

\reference{} Balogh, M.L., Navarro, J.F., \& Morris, S.L. 2000, ApJ, 540, 113

\reference{} Bell, E.F., Wolf, C., Meisenheimer, K., Rix, H.-W., Borch, A., Dye, S., Kleineinrich, M.,  Wisotzki, L.,
\&  McIntosh, D. 2004,  ApJ, 608, 752

\reference{} Bertin, E., Arnouts, S. 1996, A\&AS, 117, 393

\reference{} Blakeslee, J. et al. 2003, ApJ, 596, 143

\reference{} Blakeslee, J. et al. 2006, ApJ, 644, 30

\reference{} Bond, J.R., Cole, S., Efstathiou, G., \& Kaiser, N., 1991, ApJ, 379, 440

\reference{} Bower, R., Kodama, T., Terlevich, A. 1998, MNRAS, 299, 1193

\reference{} Bruzual, G., Charlot, S. 2003, MNRAS, 344, 1000

\reference{} Cavaliere, A., Colafrancesco, S., \& Menci, N., 1992, ApJ, 392, 41

\reference{} Cavaliere, A.,  \& Vittorini, V.  2000, ApJ, 543, 599

\reference{} Ciotti, L. \& Ostriker, J.P. 1997, ApJ, 487, L105

\reference{} Cooper, M.C. et al. 2006, MNRAS, 370, 198

\reference{} Cooper, M.C. et al. 2007, MNRAS, 370, 198

\reference{} Cowie, L.L., Songaila, A., Hu, E.M., Cohen, J.G. 1996, AJ, 112, 839

\reference{} Dekel, A \& Birnboim, Y. 2006, MNRAS, 368, 2

\reference{} De Lucia, G. et al. 2007, MNRAS, 374, 809

\reference{} Demarco, R. et al. 2007, ApJ, 663, 164

\reference{} Diaferio, A., Kauffmann, G., Balogh, M.L., White, S.D.M., Schade, D., Ellingson, E. 2001, MNRAS, 323, 999

\reference{} Di Matteo, T., Springel, V., Hernquist, L., 2005, Nature, 433, 604

\reference{} Faber, S.M. et ak. 2007, ApJ, 665, 265

\reference{} Fabian, A. 1999, MNRAS, 308, 39

\reference{} Gerke, B.F. et al. 2007, MNRAS, 376, 1425
ApJ, 622, 116

\reference{} Gobat, R. et al. 2008, A\&A, submitted

\reference{} Haehnelt, M.J., Natarajan, P., \& Rees, M.J., 1998, MNRAS, 300, 817

\reference{} Homeier, N.L. et al. 2006, ApJ, 647, 256

\reference{} Hopkins A. M., 2004, ApJ, 615, 209

\reference{} Kauffmann, G. 1996, MNRAS, 281, 487

\reference{} Kodama, T., Arimoto, N. 1997, A\&A, 320, 41

\reference{} Lacey, C.G. and Cole, S. 1993, MNRAS,  262, 627

\reference{} Lapi, A., Cavaliere, A., \&  Menci, N. 2005, ApJ, 619, 60

\reference{} Larson, R.B., Tinsley, B.M., \& Caldwell, C.N. 1980, ApJ, 237, 692

\reference{} Mei, S. et al. 2006a, ApJ, 639, 81

\reference{} Mei, S. et al. 2006b, ApJ, 644, 759

\reference{} Mei, S. et al. 2008, ApJ, submitted

\reference{} Menci, N., Cavaliere, A., Fontana, A., Giallongo, E., Poli, F.  2002,
ApJ, 578, 18

\reference{} Menci, N., Cavaliere, A., Fontana, A., Giallongo, E., Poli, F., Vittorini, V. 2003, ApJ, 587, L63

\reference{} Menci, N., Cavaliere, A., Fontana, A., Giallongo, E., \& Poli, F. 2004a,
ApJ, 604, 12

\reference{} Menci, N., Fiore, F., Perola, G.C., Cavaliere, A. 2004b, ApJ, 606, 58

\reference{} Menci, N., Fontana, A., Giallongo, E., \& Salimbeni, S. 2005,
ApJ,  632, 49

\reference{} Menci, N., Fontana, A., Giallongo, E., Grazian, A. \& Salimbeni, S. 2006,
ApJ,  647, 653

\reference{} Neistein, E., van den Bosch, F.C., Dekel, A. 2006, MNRAS, 372, 933

\reference{} Rettura, A. et al. 2006, A\&A, 458, 717 

\reference{} Rettura, A. et al. 2008, submitted 

\reference{} Romeo, A.D., Napolitano, N.R., Covone, G., 
Sommer-Larsen, J., Antonuccio-Delogu, V., Capaccioli, M. 2008, preprint
(astro-ph/0804.1517)

\reference{} Silk, J. \& Rees, M.J. 1998, A\&A, 331, 1

\reference{} Spitzer, L, Jr., Baade, W. 1951, ApJ, 113, 413

\reference{} Strateva, I. et al. 2001, AJ, 122, 1861-1874

\reference{} Strazzullo, V. et al. 2006, A\&A, 450, 909

\reference{} Vanzella, E. et al. 2008, A\&A,  478, 83

\reference{} Weinmann, S.M., van der Bosch, F.C., Yang X., Mo, H.j. 2006, MNRAS, 366, 2

\end{references}
\end{document}